\documentclass[preprint,prc,showpacs,preprintnumbers,amsmath,amssymb,floatfix]{revtex4}
\usepackage{graphicx,latexsym,amssymb, epsfig}
\usepackage{multirow,amsmath,array,booktabs}
\usepackage{subfigure}
\usepackage{color}
\usepackage{bm}
\usepackage[figuresright]{rotating}
\usepackage[section]{placeins}
\usepackage{slashed}
\usepackage{graphicx}
\usepackage{tikz}
\usepackage{pdfpages}

\newcommand{\nc}{\newcommand}       
\nc{\vc}[1] {\mbox{\boldmath $#1$}} 
\nc{\del}       {\partial}              
\nc{\bra}       {\langle}               
\nc{\ket}       {\rangle}               
\nc{\bras}[1]   {\langle #1|}           
\nc{\kets}[1]   {|#1\rangle}            
\nc{\mapleft}[1]{           
 \smash{\mathop{\,          %
  \hbox to 1.5cm{\rightarrowfill}\, }\limits_{#1}}}
\nc{\beq}     {\begin{eqnarray}} \nc{\eeq}    {\end{eqnarray}}
\nc{\nn}      {\\\nonumber} \nc{\vs}      {\vspace{-0.275cm}}
\nc{\fra}    {\frac{1}{2}}
\nc{\mb}        {\mathbf}

\begin{document}

\title{The $\Xi N$ interaction constrained by recent $\Xi^-$ hypernuclei experiments}

\author{Jinniu Hu}
\email{hujinniu@nankai.edu.cn}
\affiliation{School of Physics, Nankai University, Tianjin 300071,  China}

\author{Ying Zhang}
\email{yzhangjcnp@tju.edu.cn}
\affiliation{Department of Physics, Faculty of Science, Tianjin University, Tianjin 300072, China}

\author{Hong Shen}
\email{songtc@nankai.edu.cn}
\affiliation{School of Physics, Nankai University, Tianjin 300071,  China}





\date{\today}
\begin{abstract}
	The $\Xi N$ interaction is investigated in the quark mean-field (QMF) model based on recent observables of the $\Xi^-+^{14}\rm{N}$ ($_{\Xi^-}^{15}\rm{C}$)  system. The experimental data about the binding energy of $1p$-state  $\Xi^-$ hyperon in $_{\Xi^-}^{15}\rm{C}$ hypernuclei at KISO, IBUKI, E07-T011, E176-14-03-35 events are conflated as $B_{\Xi^-}(1p)=1.14\pm0.11$ MeV. With this constraint, the coupling strengths between the vector meson and $\Xi$ hyperon are fixed in three QMF parameter sets. Meanwhile, the $\Xi^-$ binding energy  of $1s$ state in $_{\Xi^-}^{15}\rm{C}$ is predicted as $B_{\Xi^-}(1s)=5.66\pm0.38$ MeV with the same interactions, which are completely consistent with the data from the KINKA and IRRAWADDY events. Finally, the single $\Xi N$ potential is calculated in the symmetric nuclear matter in the framework of QMF models. It is $U_{\Xi N}=-11.96\pm 0.85$ MeV at nuclear saturation density, which will contribute to the study on the strangeness degree of freedom in compact star.
\end{abstract}


\maketitle

\section{Introduction}
The strangeness is a very important degree of freedom in nuclear physics, which arises in the hypernuclei and the core of compact stars with increments of energy and density. The hyperon including strange quarks is not eliminated by the Pauli principle in finite nuclei. Therefore, it is a very powerful probe to detect the information of strong interaction in nuclear medium, which is also essential to the equation of state of the compact star at high-density region~\cite{gal16,oertel17}. The ordinary strangeness nuclear system is the single-$\Lambda$ hypernucleus with one strange quark, $s$, i.e., $S=-1$. There have been many observables about single-$\Lambda$ hypernuclei  from light to heavy mass region~\cite{hashimoto06,feliciello15}, where the $\Lambda N$ interaction is well extracted from the single-$\Lambda$ binding energy, $B_\Lambda$, in various theoretical methods,~\cite{hiyama08,wirth14,millener08,li13,schulze13,cui15,zhou16,shen06,liu18,saito07}.

For the hypernuclei including two strange quarks, i.e., double $\Lambda$ hypernuclei or $\Xi$ hypernuclei,  the relevant experiments are much fewer than those about single-$\Lambda$ hypernuclei, since the lifetime of $\Xi$ hyperon is shorter than the one of $\Lambda$ hyperon. There are only a few events about the double $\Lambda$ hypernuclei in light mass region~\cite{ahn13,hiyama18}. With the developments of experimental technique and analysis method, such as ``emulsion-counter hybrid method" and ``overall scanning method"~\cite{yoshida17}, a deeply bound $\Xi^-$ hypernucleus was confirmed in the reaction, $\Xi^-+^{14}\rm{N}\rightarrow ^{14}_{\Lambda}\rm{Be}+ ^{5}_{\Lambda}\rm{He}$ in KEK E373 emulsion, which was also named as KISO event~\cite{nakazawa15}. It provides strong evidence that the $\Xi N$ interaction should be attractive.

In the KISO event, the single-$\Xi^-$ binding energy, $B_{\Xi^-}$ cannot be well determined, which is $B_{\Xi^-}=3.87\pm0.21$ MeV or $B_{\Xi^-}=1.03\pm0.18$ MeV~\cite{hiyama18}. Therefore, many theoretical investigations were done to estimate whether the $\Xi^-$ occupies on $1s$ state or $1p$ state in KISO event, such as relativistic mean-field (RMF) model, Skyrme-Hartree-Fock (SHF) model~\cite{sun16}, and quark meson coupling (QMC) model~\cite{shyam19}. We also adopted the quark mean-field (QMF) model to calculate the $_{\Xi^-}^{15}\rm{C}$ hypernucleus~\cite{hu17}, where the baryons are consisting of three constituent quarks and they interact with each other through exchanging the mesons. It was found that the $\Xi^-$ hyperon should be $1p$ state in KISO event by employing the results from cluster models on  $_{\Xi^-}^{12}\rm{Be}$, when the $\Xi N$ interaction at nuclear saturation density was fixed at $U_{\Xi N}(\rho_0)=-12$ MeV. Furthermore, it was also extracted through the Wood-Saxon potential by fitting the single $\Xi^-$ binding energy of $\Xi^-$ hypernuclei~\cite{friedman21}. Meanwhile, the in-medium $\Xi N$ potential was calculated with the realistic $\Xi N$ interaction generated by the lattice QCD simulation, chiral effective field potential, and Nijmegen ESC16 potential in the framework of Brueckner-Hartree-Fock model, where $U_{\Xi N}(\rho_0)$ have very large uncertainties~\cite{sasaki20,haidenbauer19,kohno19,nagels20}. It was about $-24,~-10$, and $-5$ MeV in Wood-Saxon potential, lattice QCD simulation, and chiral potential, respectively.

Recently, more events about $\Xi^-+^{14}\rm{N}$ system were analyzed from the KEK E373 and J-PARC E7 experiments, such as IBUKI event, KINKA event, IRRAWADDY event, and E07-T011 event~\cite{hayakawa21,yoshimoto21}. The experimental data of $\Xi^-$ binding energy, $B_{\Xi^-}$ in $_{\Xi^-}^{15}\rm{C}$ is greatly enriched in its $1s$ and $1p$ states. Therefore, it is time to conflate these latest experimental results to better determine the $\Xi N$ potential in the nuclear medium.

In this work, the $1p$  $\Xi^-$ binding energy $B_{\Xi^-}(1p)$ of $_{\Xi^-}^{15}\rm{C}$ from present observations will be combined to extract the coupling strengths between mesons and $\Xi$ hyperon in the QMF model. Then the experimental information of its $1s$ state will be used to examine the validity of these interactions. Finally, the $\Xi N$ potential will be self-consistently evaluated at nuclear saturation density in nuclear matter with the same framework. 

\section{The $\Xi^-$ hypernuclei in quark mean-field model}\label{sec-QMF}
The baryons interact with each other through exchanging mesons and photons in quark mean-field (QMF) model~\cite{hu14a,hu14b,xing16,xing17}. The Lagrangian can be evaluated as,
\beq
&&{\cal L}_{\rm QMF}
=\sum_{B=N,\Xi^-}\bar\psi_B
\bigg[i\gamma_\mu\partial^\mu-M_B^*
-g_{\omega B}\omega_\mu\gamma^\mu\nn
&&+\frac{f_{\omega B}}{2M_B}\sigma^{\mu\nu}\partial_\nu\omega_\mu
-g_{\rho B} \gamma^\mu\vec\tau_B\cdot\vec\rho_\mu 
-e\frac{q_B(1-\tau_{B,3})}{2}\gamma^\mu A_\mu \bigg]\psi_B\nn
&&
+\frac{1}{2} \partial^\mu\sigma\partial_\mu\sigma
-\frac{1}{2} m_\sigma^2\sigma^2
-\frac{1}{3} g_2\sigma^3
-\frac{1}{4} g_3\sigma^4\nn
&&
-\frac{1}{4} W^{\mu\nu}W_{\mu\nu}
+\frac{1}{2} m_\omega^2\omega^2
+\frac{1}{4} c_3\omega^4\nn
&&
-\frac{1}{4} \vec R^{\mu\nu}\vec R_{\mu\nu}
+\frac{1}{2} m_\rho^2\rho^2
-\frac{1}{4}F^{\mu\nu}F_{\mu\nu},
\eeq
where $\psi_B$ is the field operator of baryon. $\sigma,~\omega_\mu$, and $\vec\rho_\mu$ represent the isoscalar-scalar, isoscalar-vector, and isovector-vector mesons, respectively. The isospin vectors are denoted as arrows. Meanwhile, the tensor coupling term between $\omega$ meson and baryon, $\frac{f_{\omega B}}{2M_B}\sigma^{\mu\nu}\partial_\nu\omega_\mu$ is only considered for $\Xi^-$ hyperon to reduce the spin-orbit splitting of hypernuclei. The $q_B$ is the charge of baryon as $q_B=e$ for proton and $q_B=-e$ for $\Xi^-$ hyperon. The third isospin component is given by $\tau_{B,3}$, which both are $-1$ for proton and $\Xi^-$ hyperon and three tensor operators about the vector and the photon fields, $W^{\mu\nu},~\vec R^{\mu\nu}$, and $F^{\mu\nu}$ are shown as follows,
\beq
W^{\mu\nu}&=&\partial^\mu\omega^\nu-\partial^\nu\omega^\mu,\nn
\vec R^{\mu\nu}&=&\partial^\mu\vec\rho^\nu-\partial^\nu\vec\rho^\mu,\nn
F^{\mu\nu}&=&\partial^\mu A^\nu-\partial^\nu A^\mu.
\eeq

The equations of motion of baryons and mesons can be generated by the Euler-Lagrange equation. The no-sea and mean-field approximations must be taken into account to achieve the numerical calculations. Furthermore, the spatial terms of vector fields vanish in spherical symmetry nuclei. Therefore, the $\omega,~\rho$, and $A$ are adopted to indicate their time components. Finally, these equations can be derived as
\beq\label{eom}
&&\bigg[i\gamma_{\mu}\partial^{\mu}-M_B^*
-g_{\omega B}\omega\gamma^0
+\frac{f_{\omega B}}{2M_B}\sigma^{0i}\partial_i\omega\nn
&&-g_{\rho B}\rho\tau_{B,3}\gamma^0-q_B\frac{(\tau_{B,3}-1)}{2}A\gamma^0\bigg]\psi_B
=0,\nn
&&-\Delta\sigma+m_\sigma^2\sigma+g_2\sigma^2+g_3\sigma^3
=\sum_{B=N,\Xi^-}-\frac{\partial M_B^*}{\partial\sigma}
\langle\bar\psi_B\psi_B\rangle,\nn
&&-\Delta\omega+m_\omega^2\omega+c_3 \omega^3\nn
&&=\sum_{B=N,\Xi^-}
g_{\omega B}\langle\bar\psi_B\gamma^0\psi_B\rangle
-\frac{f_{\omega\Xi}}{2M_B}
\partial_i\langle\bar\psi_B\sigma^{0i}
\psi_B\rangle,\nn
&&-\Delta\rho+m_\rho^2\rho=\sum_{B=N,\Xi^-}
g_{\rho B}\langle\bar\psi_B\tau_{B,3}\gamma^0\psi_B\rangle,\nn
&&-\Delta A=\sum_{B=N,\Xi^-}
q_B\langle\bar\psi_B\frac{(1-\tau_{B,3})}{2}\gamma^0\psi_B\rangle.\nn
\eeq

In the QMF model, the effective baryon mass, $M^*_B(\sigma)$ is determined by the quark model, where the baryons consist of three constituent quarks. They are confined by central  harmonics confinement potentials, $U(r)=\frac{1}{2}(1+\gamma^0)(a_qr^2+V_q)$. Therefore, the baryon mass is provided by the dynamics mechanism of quarks, i.e., their eigen energies, center of mass corrections, pion contribution, and gluon effect. The other coupling constants between mesons and baryons have been fixed by fitting the ground state properties of several double-magic nuclei~\cite{xing16}. The only free parameter for the $\Xi^-$ hypernuclei in QMF model is the strength between $\omega$ meson and $\Xi^-$ hyperon, $g_{\omega\Xi}$, which is also related to the single $\Xi N$ potential,
\beq\label{xipot}
U_{\Xi N}=-\frac{\partial M_\Xi^*}{\partial\sigma}\sigma+g_{\omega\Xi}\omega.
\eeq

The equations of motion of baryons and mesons in Eq.~(\ref{eom}) can be self-consistently solved with the numerical scheme and generate the ground-state properties of $\Xi^-$ hypernucleus such as its total binding energy, single-particle energy levels, and so on after treating the center-of-mass corrections of many-particle system with the microscopic method and the pairing correlations of open-shell nuclei with BCS theory.

\section{The numerical results and discussions}
In the past years, the information about $\Xi N$ potential was very unclear due to various challenges in the experiments of $\Xi$ hypernuclei. Recently, several events about the  $_{\Xi^-}^{15}\rm{C}$ hypernuclei in KEK E373 and J-PARC E07 experiments were analyzed~\cite{yoshimoto21}. Most of $\Xi^-$ hyperon in  $_{\Xi^-}^{15}\rm{C}$  possibly occupied $1s$ or $1p$ state in these events. The first certain deep bound $\Xi^-$ hypernuclei was detected in the KISO event of  KEK-E373 experiment, $\Xi^-+^{14}\rm{N}\rightarrow ^{14}_{\Lambda}\rm{Be}+ ^{5}_{\Lambda}\rm{He}$, where the binding energy of $\Xi^-$, $B_{\Xi^-}$, was $3.87\pm0.21$ or $1.03\pm 0.18$ MeV. The $\Xi^-$ hyperon may stay in $1s$ or $1p$ state~\cite{nakazawa15}. However, many theoretical works concluded that it is most probably the $1p$ state~\cite{sun16,hu17,shyam19}. Later, another double strangeness hypernucleus event was measured in the J-PARC E07 experiment, named as the IBUKI event, the $B_{\Xi^-}$ is $1.27\pm0.21$ MeV, which was interpreted as a $1p$ state~\cite{hayakawa21}. Furthermore, there are another two events, 14-03-35 in early experiment KEK176~\cite{aoki09} and T011 in J-PARC E07 related to $_{\Xi^-}^{15}\rm{C}$, which provided the values of $B_{\Xi^-}$ at $1p$ state. All these experimental data are listed in Table~\ref{tab1}.
\begin{table}[htb]
	\centering
	\caption{The available single-$\Xi^-$ binding energies at $1p$ state in $_{\Xi^-}^{15}\rm{C}$  from various experiments.}
	\label{tab1}
	\begin{tabular}{c c c}
		\hline
		\hline
		Experiment&Event&$B_{\Xi^-}(1p)$ (MeV)\\
		\hline		
	    KEK E176~\cite{aoki09}&14-03-35 & $1.18\pm0.22$    \\
		KEK E373~\cite{nakazawa15}&KISO&$1.03\pm0.18$         \\
		J-PARC E07~\cite{hayakawa21}&IBUKI &$1.27\pm0.21$           \\
		J-PARC E07~\cite{yoshimoto21}&T011&$0.90\pm0.62$           \\
		\hline
		\hline
	\end{tabular}
\end{table}

Here, the KISO event was regarded as a $1p$  $\Xi^-$ hypernucleus. These manifold experimental results are not appropriate to determine the coupling constants $g_{\omega\Xi}$ in the QMF model. Therefore, they should be summarized at first. To consolidate these data from several independent experiments, which measure the same physical quantity, $B_{\Xi^-}(1p)$, a mathematical method named conflation will be used~\cite{hill11a,hill11b}. All observables of  $\Xi^-$ binding energy from the above events are assumed to satisfy the normal distributions, 
\beq
f(x)=\frac{1}{\sigma_i\sqrt{2\pi}}\exp\left[-\frac{(x-m_i)^2}{2\sigma^2_i}\right],
\eeq
where $m_i$ is the mean value, and the $\sigma_i$ represents the standard deviation. It can be proved that the best linear unbiased estimation for many independent observations with normal distributions is still a normal distribution in conflation method with 
\beq
m_t&=&\sum^n_i m_i\sigma_i^{-2}\left(\sum^n_i \sigma_i^{-2}\right)^{-1},\nn
\sigma_t&=&\left(\sum^n_i \sigma_i^{-2}\right)^{-1/2}.
\eeq

\begin{figure}[htb]
	\centering
	\includegraphics[width=0.6\textwidth, angle=0]{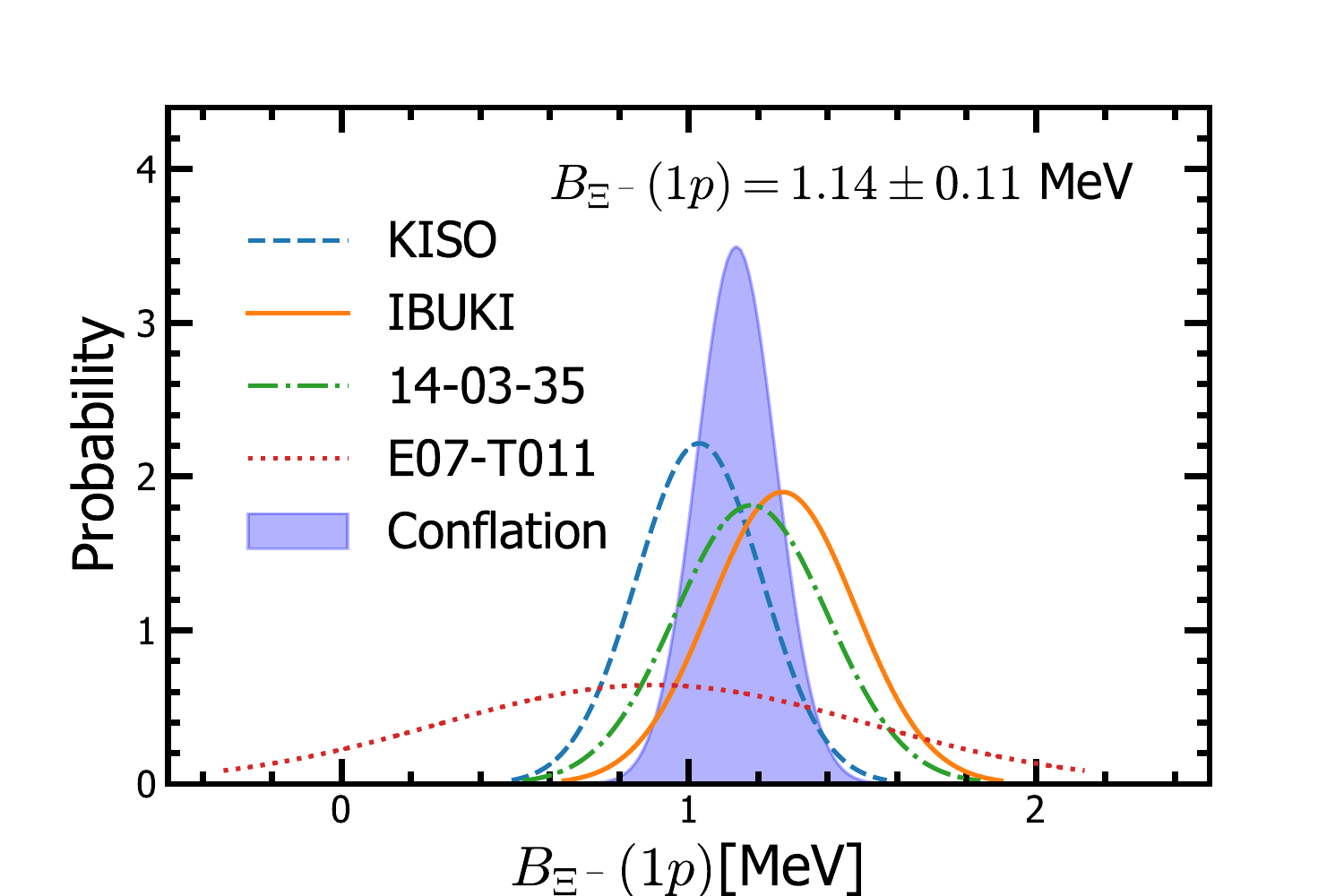}
	\caption{The normal distributions of $\Xi^-$ binding energy at $1p$ state, $B_{\Xi^-}(1p)$ of $_{\Xi^-}^{15}\rm{C}$  in 14-03-35, KISO, IBUKI, and T011 events and their conflation distribution. }\label{xi1p}
\end{figure}
In Fig.~\ref{xi1p}, the probabilities of $B_{\Xi^-}$ at $1p$ state in  $_{\Xi^-}^{15}\rm{C}$ from 14-03-35, KISO, IBUKI, and T011 events are shown, when the observables from KEK E176, E373, and J-PARC E07 were considered as the normal distributions. Their conflating distribution is given as the shadow region. It can be found that the conflating mean value is close to the means of four independent experiments with a smaller variance, which indicates a better accuracy since the total valid measurements are increasing compared to the single experimental data. Therefore, the $\Xi^-$ binding energy at $1p$ state in $_{\Xi^-}^{15}\rm{C}$  can be well determined as $B_{\Xi^-}=1.14\pm0.11$ MeV within present conflation method.

Once the values of $B_{\Xi^-}(1p)$ were constrained, the only free parameter in the QMF model for $\Xi^-$ hypernuclei, $g_{\omega\Xi }$ will be fixed. The other coupling constants between mesons and baryons and the strengths of confinement potentials have been determined by the ground-state properties of finite nuclei and free baryon masses, in our previous works~\cite{xing16,xing17}. Furthermore to investigate the influences of quark masses, these parameter were separated into  three sets, i.e., QMF-NK1, QMF-NK2, and QMF-NK3, where the constituent quark masses were adopted as $m_u= 250, ~m_s=330$ MeV;  $m_u= 300, ~m_s=380$ MeV; $m_u= 350, ~m_s=430$ MeV, respectively for $u$ and $s$ quarks.  Based on these parameter sets, the $g_{\omega\Xi }/g_{\omega N}$ can be generated  by the $1p$ binding energy of $_{\Xi^-}^{15}\rm{C}$,  $B_{\Xi^-}=1.14\pm0.11$ MeV. We choose its two boundary values, $-1.03$ MeV and 
$-1.25$ MeV to obtain two $g_{\omega\Xi }$ for each  of three parameter sets, named as QMF-NK1X1, QMF-NK1X2, and so on. Their values are listed in Table~\ref{tab2}. Here, the tensor coupling term between $\omega$ and $\Xi^-$ is introduced with $f_{\omega\Xi}=-0.4g_{\omega\Xi}$  so that the spin-orbit splitting of $\Xi^-$ hypernuclei is largely reduced. In the present work, the $1p_{3/2}$ state of $\Xi^-$ hyperon will be used to denote the $1p$ state in experimental measurements.

\begin{table}[htb]
	\centering
	\caption{The coupling constants between $\omega$ meson and $\Xi$ hyperon in QMF model with different quark masses in terms of the conflating constraints of $1p$ $\Xi^-$ binding energies of $_{\Xi^-}^{15}\rm{C}$ and their predictions on its $1s$ state binding energies $B_{\Xi^-} (1s)$. The binding energies are in unit of MeV.}
	\label{tab2}
	\begin{tabular}{c c c c}
		\hline
		\hline
		Sets&$g_{\omega\Xi}/g_{\omega N}$&$B_{\Xi^-}(1p)$&$B_{\Xi^-}(1s)$ \\
		\hline		
		QMF-NK1X1&$0.5024$& $-1.03$&$-5.19$    \\
		QMF-NK1X2&$0.4954$& $-1.25$&$-5.91$    \\
		QMF-NK2X1&$0.4832$& $-1.03$&$-5.27$    \\
		QMF-NK2X2&$0.4771$& $-1.25$& $-6.01$    \\
		QMF-NK3X1&$0.4694$& $-1.03$&$-5.41$    \\
		QMF-NK3X2&$0.4638$& $-1.25$&$-6.11$    \\
		\hline
		\hline
	\end{tabular}
\end{table}

Meanwhile, the $\Xi^-$ binding energy at $1s$ state of $_{\Xi^-}^{15}\rm{C}$  can be predicted with the same parameters. Their corresponding values are also tabulated in the last column of Table~\ref{tab2} about $-5.19$--- $-6.11$ MeV. Because all of these $B_{\Xi^-}(1s)$ is produced by assuming that the $B_{\Xi^-}(1p)$ from experimental data satisfies a normal distribution, it is natural to analyze them with the same distribution and calculate their mean value and standard deviation. Finally, the $B_{\Xi^-}(1s)$ can be predicated as $5.66\pm0.38$ MeV in the QMF model. At the experimental aspect, there is only a certain event, IRRAWADDY at J-PARC E07 experiment, which pointed the $B_{\Xi^-}(1s)=6.27\pm0.27$ MeV in $_{\Xi^-}^{15}\rm{C}$. Our results are consistent with the IRRAWADDY event quite well. Furthermore, in another event, KINKA in the KEK E373 experiment, the $\Xi^-$ binding energy was not fixed well, which can be $4.96\pm0.77$ or $8.00\pm0.77$ MeV. According to the present calculations, it should be $4.96\pm0.77$ MeV and be interpreted as a $1s$ state.
\begin{figure}[htb]
	\centering
	\includegraphics[width=0.6\textwidth, angle=0]{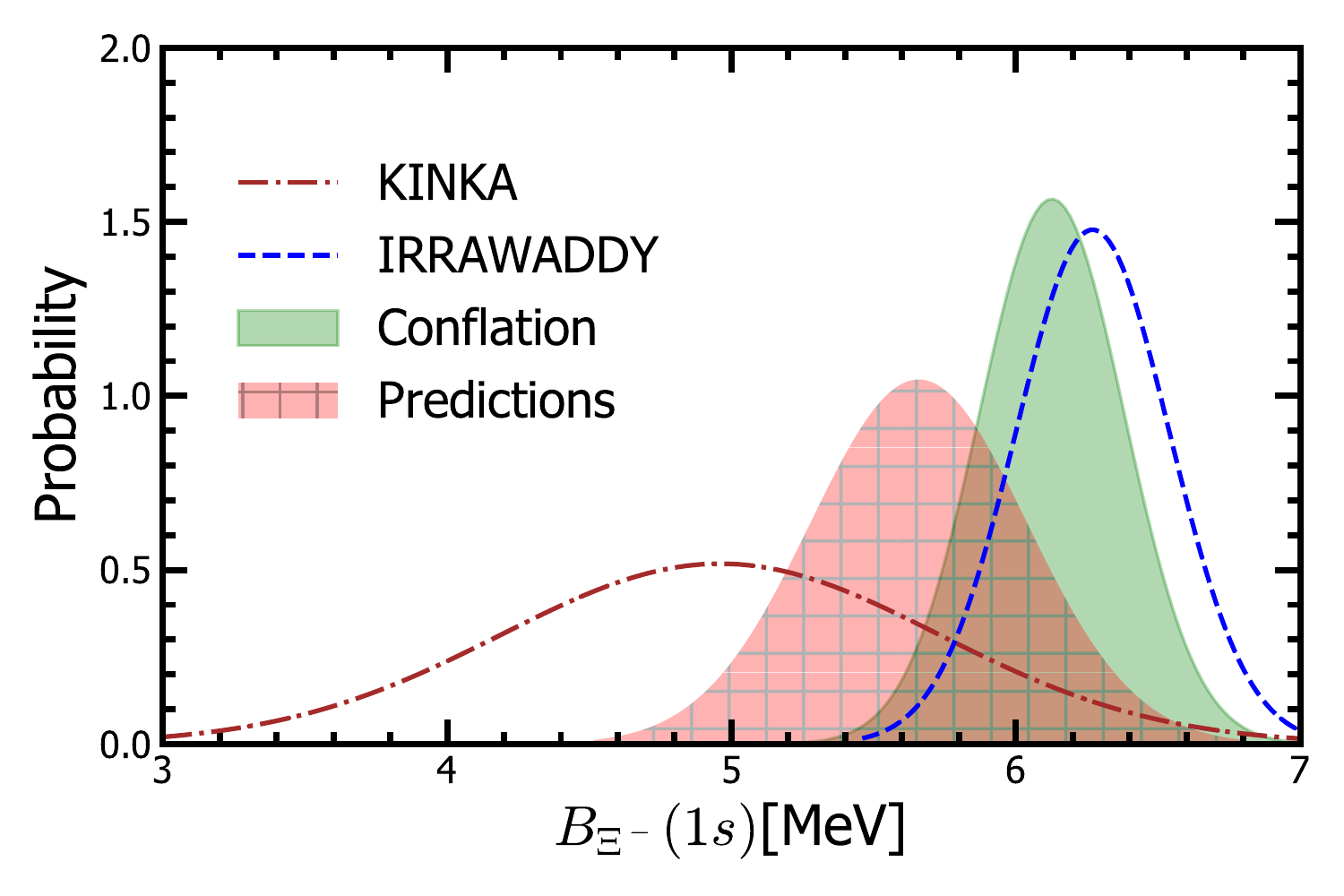}
	\caption{The distributions of $\Xi^-$ binding energy at $1s$ state, $B_{\Xi^-}(1s)$ of $_{\Xi^-}^{15}\rm{C}$  in IRRAWADDY and KINKA events, their conflation distribution and the prediction distribution from QMF model. }\label{xi1s}
\end{figure}

The probabilities of $\Xi^-$ binding energy at $1s$ state from IRRAWADDY and KINKA events are plotted in Fig.~\ref{xi1s}, where the KINKA is considered as $1s$ state with  $B_{\Xi^-}(1s)=4.96\pm0.77$ MeV.  In this assumption, their conflation value is $6.13\pm0.25$ MeV. It has a very wide overlap with the prediction from the present QMF model,  $5.66\pm0.38$ MeV.

On the other hand, the magnitude of $\Xi N$ potential in the nuclear matter is also very important for the discussions of hyperon in  the neutron star, which is strongly correlated to the onset densities of $\Xi$ hyperons in the core region of the neutron star. With the Eq.~(\ref{xipot}), the single $\Xi N$ potential in the symmetric nuclear matter is easily evaluated, where the $\sigma$ and $\omega$ fields are obtained through solving the equations of motion of nucleons and mesons in nuclear matter.

\begin{figure}[htb]
	\centering
	\includegraphics[width=0.6\textwidth, angle=0]{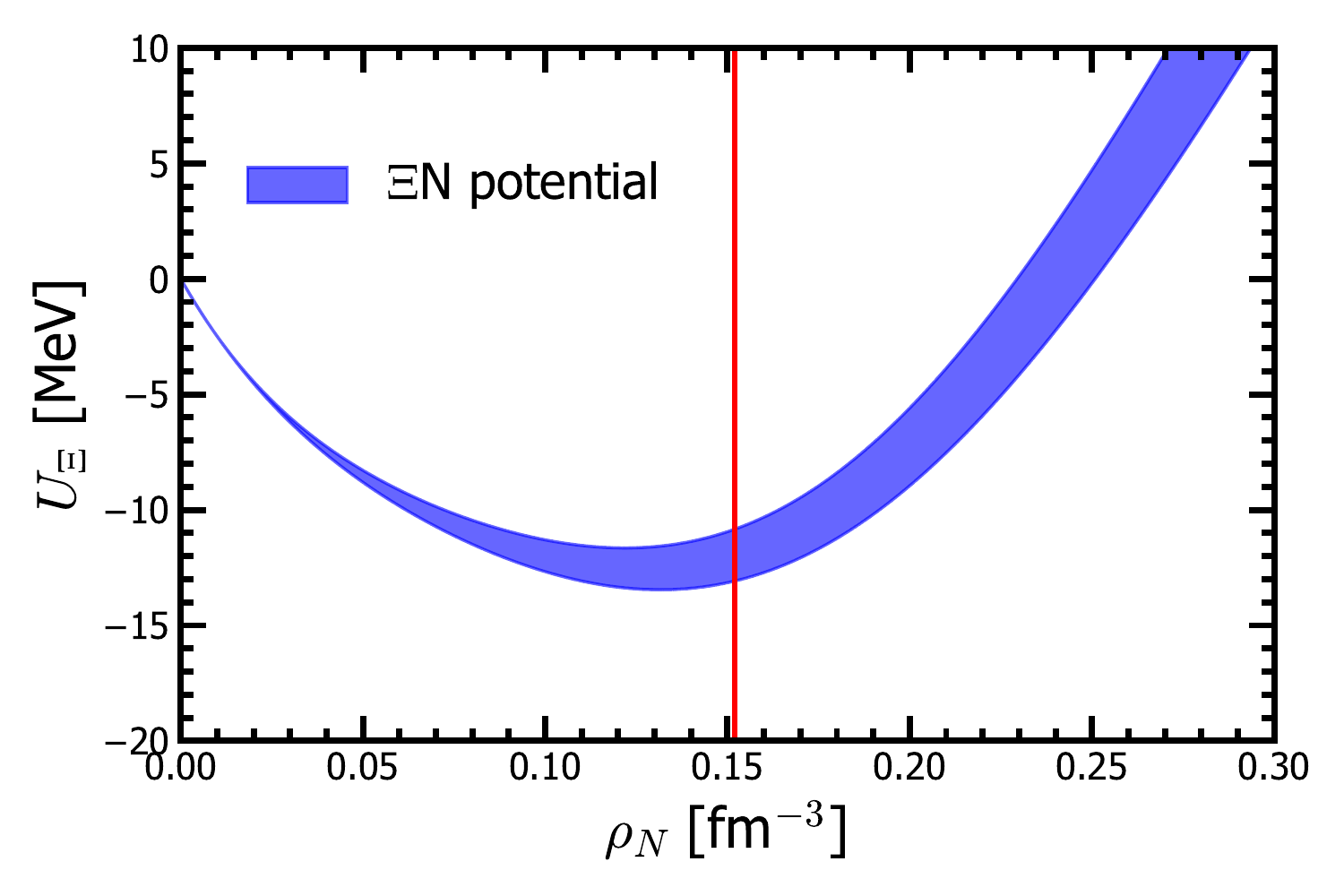}
	\caption{The single $\Xi N$ potentials in symmetric nuclear matter from QMF models as functions of nuclear density.}\label{pot}
\end{figure}

The $\Xi N$ potentials as functions of nuclear density are given in Fig.~\ref{pot} with different QMF parameter sets. It is an attractive potential and decreases in low-density region. It becomes saturated around $\rho_N=0.14$ fm$^{-3}$. In the high-density region, it rapidly increases and inverses to a repulsive potential. The nuclear saturation densities in QMF-NK1, QMF-NK2, and QMF-NK3 are around $0.152$ fm$^{-3}$~\cite{xing16}. At this density, the $\Xi N$ potentials are around $-10.72$---$-13.15$ MeV.

\begin{figure}[htb]
	\centering
	\includegraphics[width=0.6\textwidth, angle=0]{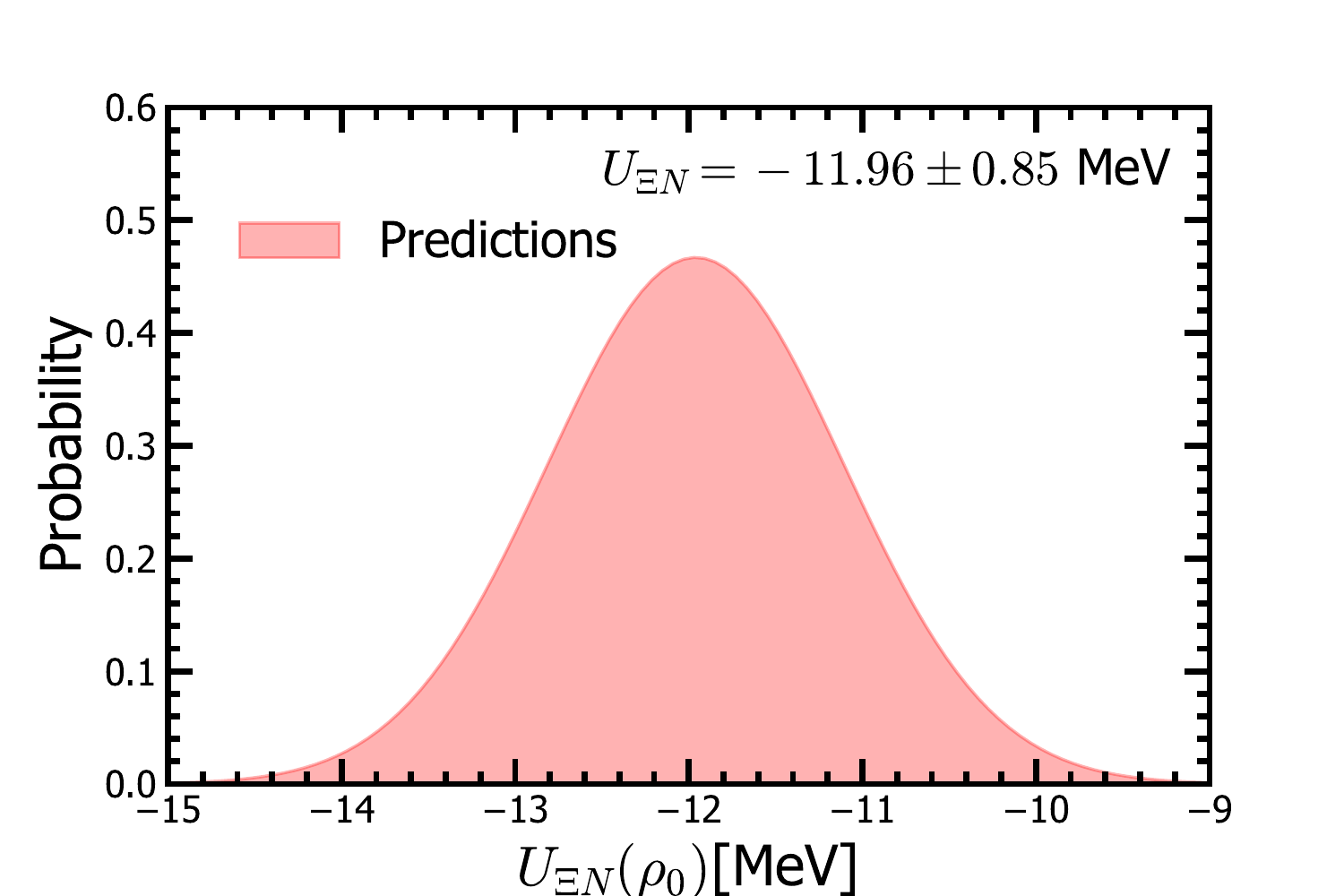}
	\caption{The distribution of $\Xi N$ potential at nuclear saturation density from the QMF models.}\label{potdis}
\end{figure}

The normal distribution of $\Xi N$ potential, $U_\Xi$ at nuclear saturation density $\rho_0$ from QMF models is obtained with the same scheme about the $B_{\Xi^-}$ of $1s$ state $\Xi^-$ hyperon in $_{\Xi^-}^{15}\rm{C}$ , which is shown in Fig.~\ref{potdis} and $U_\Xi(\rho_0)=-11.96\pm 0.85$ MeV. The magnitude of this potential is obviously smaller than the recent analysis with the Wood-Saxon potential, where the $\Xi N$ potential depth is about $24.3\pm0.8$ MeV~\cite{friedman21}. Therefore, this Wood-Saxon potential also provided a more bound state of $1s$ $\Xi^-$ with $10$ MeV, which is not consistent with the results from the KINKA and IRRAWADDY events. Furthermore, the  $\Xi N$ potential was also discussed with chiral NLO potential with $G$-matrix, where $U_\Xi(\rho_0)\sim-5$---$-8$ MeV~\cite{haidenbauer19,kohno19}. Its prediction about $\Xi^-$ binding energy was a little bit smaller than the present experiment constraints. The $\Xi N$ potential at nuclear saturation density was also calculated with the HAL-QCD $\Xi N$ potential, which was about $-10.6$---$-16.2$ MeV with $G$-matrix method~\cite{nagels20}.  It may generate the lightest $\Xi$ hypernucleus with the same lattice potential in the framework of the Gaussian expansion method~\cite{hiyama20}.

\section{Summary and outlook}\label{sec-concl}
The $\Xi N$ potential was studied in the framework of quark-mean field (QMF) model, with the recent experimental constraints about the $\Xi^-+ ^{14}\rm{N}$ system. Firstly, several binding energies of $1p$ state $\Xi^-$ hyperon in $_{\Xi^-}^{15}\rm{C}$ hypernucleus from the KEK E176, KEK E373, and J-PARC E07 experiments were conflating as $B_{\Xi^-}=1.14\pm0.11$ MeV. With this data, the coupling constant between $\omega$ meson and $\Xi$ hyperon was fixed in the case of different quark masses, while the other parameters in the QMF model have been determined by the free baryon masses and the ground-state properties of doubly magic nuclei. After that, the binding energy of $1s$ state $\Xi^-$ hyperon in $_{\Xi^-}^{15}\rm{C}$ was predicted as $B_{\Xi^-}(1s)=5.66\pm0.38$ MeV, which is consistent with the observables from the IRRAWADDY event. Besides, the $\Xi^-$ hyperon in the KINKA event can be interpreted as a $1s$ state with present results, whose binding energy should be $4.96\pm0.77$ MeV. 

The single $\Xi N$ potential in the symmetric nuclear matter was calculated with the same parameter sets. It has a strong attractive contribution below $0.25$ fm$^{-3}$. Its magnitude at nuclear saturation density is $-11.96\pm0.85$ MeV from present QMF models, which accords to the results of HAL-QCD potential within the Brueckner-Hartree-Fock model and is smaller than the analysis of Wood-Saxon potential. It will be much helpful to the investigations of the neutron star to better discuss the strangeness degree of freedom in compact star.

Although there has been strong evidence to show that the $\Xi N$ potential provides an attractive contribution with recent experiments in $_{\Xi^-}^{15}\rm{C}$, this experimental information still has a large uncertainty. More events about the $\Xi$ hypernuclei are expected to understand the $\Xi N$ potential in the nuclear medium better.

J. Hu would like to thank Dr. Minggang Zhao for his useful knowledge about the experimental data analysis. This work was supported in part by the National Natural Science Foundation of China (Grants  No. 11775119) and the Natural Science Foundation of Tianjin.

\end{document}